\documentclass[aps,pra,floatfix,a12]{revtex4}
\usepackage{amsfonts}
\usepackage{amsmath}
\usepackage{amsthm}
\usepackage{amssymb}
\usepackage{graphicx}
\usepackage{appendix}
\usepackage{hyperref}
\usepackage{textcomp}
\usepackage{multirow}
\usepackage{color}
\usepackage{bm}
\usepackage{braket}
\usepackage{subfigure}

\begin{document}

\title{Self-accelerating solitons}
\author{Boris A. Malomed$^{1,2}$}
\address{$^{1}$Department of Physical Electronics, School of Electrical Engineering, Faculty of Engineering, and Center for Light-Matter Interaction, Tel Aviv University, Tel Aviv 69978, Israel\\
$^{2}$Instituto de Alta Investigaci\'{o}n, Universidad de Tarapac\'{a}, Casilla 7D, Arica, Chile}

\begin{abstract}
Basic models which give rise to one- and two-dimensional (1D and 2D)
solitons, such as the Gross-Pitaevskii (GP)\ equations for BEC, feature the
Galilean invariance, which makes it possible to generate families of moving
solitons from quiescent ones. A challenging problem is to find models
admitting stable self-accelerating (SA) motion of solitons. SA modes are
known in linear systems in the form of Airy waves, but they are poorly
localized states. This brief review presents two-component BEC models which
make it possible to predict SA solitons. In one system, a pair of
interacting 1D solitons with opposite signs of the effective mass is created
in a binary BEC trapped in an optical-lattice potential. In that case,
opposite interaction forces, acting on the solitons with positive and
negative masses, produce equal accelerations, while the total momentum is
conserved. The second model is based on a system of GP equations for two
atomic components, which are resonantly coupled by a microwave field. The
latter model produces an exact transformation to an accelerating references
frame, thus predicting 1D and 2D stable SA solitons, including vortex rings.
\end{abstract}

\maketitle

\textbf{Introduction}. -- A basic property of one- and two-dimensional (1D
and 2D) equations which produce solitons is the Galilean invariance, which
generates solitons moving with an arbitrary velocity from a quiescent one. A
challenging issue is to construct localized states moving at a constant
acceleration, rather than constant velocity \cite{Belanger,Parker}. A
well-known fact is that the linear Schr\"{o}dinger equation, $%
iu_{z}+(1/2)u_{xx}=0$ (it is written in the form of the paraxial propagation
equation in optics, with propagation distance $z$ and transverse coordinate $%
x$) admits self-accelerating (SA) solutions in the form of Airy waves \cite%
{Berry}. Later, it was predicted \cite{Siviloglu-Christodoulides} and
experimentally demonstrated, originally in optics \cite{Siviloglu et al 2007}%
, and then in electron beams \cite{electron waves}, plasmonics \cite{Airy
plasmonics}, Bose-Einstein condensates (BECs) \cite{kli},\ acoustics \cite%
{acc}, gas discharge \cite{gas-discharge}, and water waves \cite{Airy water}
that truncated Airy waves (TAWs) can be created in these media, the
respective exact solution of the linear Schr\"{o}dinger equation being

\begin{gather}
u_{\mathrm{TAW}}\left( x,z\right) =u_{0}\mathrm{Ai}\left( \alpha x-\frac{%
\alpha ^{4}}{4}z^{2}+i\aleph \alpha ^{2}z\right) \exp \left( \aleph \alpha
x\right)  \notag \\
\times \exp \left( -\frac{i\alpha ^{6}}{12}z^{3}+\frac{i\alpha ^{3}}{2}xz-%
\frac{\aleph \alpha ^{4}}{2}z^{2}+\frac{i\aleph ^{2}\alpha ^{2}}{2}z\right) ,
\label{1D-AW}
\end{gather}%
where $\mathrm{Ai}$ is the Airy function, and constants $u_{0}$, $\alpha $,
and $\aleph >0$ define, respectively, the amplitude, internal scale, and
truncation of the Airy wave. Accordingly, the input which generates solution
(\ref{1D-AW}) is $u\left( x;z=0\right) =u_{0}\mathrm{Ai}\left( \alpha
x\right) \exp \left( \aleph \alpha x\right) $. The truncation factor is
necessary to make the total norm (power, in terms of optics) of the input
finite:
\begin{equation}
N_{\mathrm{TAW}}\equiv \int_{-\infty }^{+\infty
}|u(x)|^{2}dx=u_{0}^{2}\left( \sqrt{8\pi \aleph }\alpha \right) ^{-1}\exp
\left( 2\aleph ^{3}/3\right) ,  \label{N}
\end{equation}%
while the TAW's momentum,
\begin{equation}
P=i\int_{-\infty }^{+\infty }u_{x}^{\ast }udx,  \label{P}
\end{equation}%
is zero, in spite of the self-acceleration featured by this wave (both $N$
and $P$ are dynamical invariants of the Schr\"{o}dinger equation).

In fact, the truncation leads to gradual degradation of the TAW, as shown,
in particular, by factor $\exp \left( -\aleph \alpha ^{4}z^{2}/2\right) $ in
solution (\ref{1D-AW}). Another source of degradation is the action of
nonlinearity, as TAW is the eigenmode of the linear medium. Effects of
nonlinearity on the Airy waves were considered in many works \cite%
{Ellenbogen2}-\cite{Thawatchai}, chiefly demonstrating decay into solitons.

A possibility to create well-localized (unlike the Airy waves) SA
two-component pulses in optics was elaborated in terms of a system of
coupled nonlinear Schr\"{o}dinger (NLS)\ equations with opposite signs of
the group-velocity dispersion (GVD) in them \cite{Peschel theory}, aiming to
create two pulse components with \emph{opposite signs of their effective
masses}. In this case, the opposite interaction forces with which the
coupled components act on each other give rise to identical signs of the
acceleration. A solution for such an SA bound state was constructed
approximately, taking an unperturbed NLS soliton in the anomalous-GVD
component, and applying the Thomas-Fermi approximation to the normal-GVD
one. The so predicted optical SA pulses were demonstrated experimentally in
a roughly similar temporal-domain form, using a pair of fiber loops with
different lengths, coupled to each other at one point \cite%
{Peschel-experiment}.

It is relevant to mention that a soliton moving with a constant acceleration
(although not of the SA type) can be readily predicted taking the single NLS
equation with the self-attractive cubic term and an attractive or repulsive
local defect which moves with acceleration $a$:%
\begin{equation}
iu_{z}+(1/2)u_{xx}+|u|^{2}u=-\varepsilon \delta \left( x-az^{2}/2\right) u,
\label{delta}
\end{equation}%
where $\delta $ is the delta-function, $\varepsilon >0$ or $\varepsilon <0$
corresponding to the attractive or attractive defect, respectively. In terms
of the spatial-domain NLS equation (\ref{delta}) in optics, the
\textquotedblleft accelerating" defect represents a narrow parabolic ($%
x=az^{2}/2$) stripe in the $\left( x,z\right) $ plane, with the locally
increased ($\varepsilon >0$) or decreased ($\varepsilon <0$) value of the
refractive index. It is convenient to rewrite Eq. (\ref{delta}) in the
co-moving reference frame, applying the corresponding boost transformation
\cite{Belanger}:%
\begin{gather}
\xi \equiv x-az^{2}/2,u(x,z)\equiv v\left( \xi ,z\right) \exp \left( ia\xi
z+ia^{2}z^{3}/6\right) ,  \label{xi} \\
iv_{z}+(1/2)v_{\xi \xi }+|v|^{2}v=a\xi v-\varepsilon \delta (\xi )v.
\label{v}
\end{gather}%
In the moving frame, the soliton may be approximated by the simple
stationary solution to Eq. (\ref{v}),
\begin{equation}
v_{\mathrm{sol}}(\xi ,z)=(N/2)\mathrm{sech}\left( (N/2)\left( \xi -\xi
_{0}\right) \right) \exp \left( iN^{2}z/8\right) ,  \label{vsol}
\end{equation}%
where $N$ is the soliton's norm (see Eq. (\ref{N})), and $\xi _{0}$ is a
shift of the soliton from the position of the defect. Treating the terms on
the right-hand side of Eq. (\ref{v}) by perturbations \cite{pert}, the
soliton is considered as a quasi-particle under the action of an effective
potential,
\begin{equation}
U\left( \xi _{0}\right) =Na\xi _{0}-\left( \varepsilon N^{2}/4\right)
\mathrm{sech}^{2}\left( N\xi _{0}/2\right) .  \label{U}
\end{equation}%
This potential has an equilibrium position with sign $\mathrm{sgn}\left( \xi
_{0}\right) =-\mathrm{sgn}\left( a\varepsilon \right) $, which exists, for
given strength $\varepsilon $ of the defect, if the acceleration does not
exceed a critical value,
\begin{equation}
\left\vert a_{\max }\right\vert =|\varepsilon |N^{2}/\left( 6\sqrt{3}\right)
.  \label{amax}
\end{equation}%
The corresponding largest equilibrium value of the shift is $\left(
\left\vert \xi _{0}\right\vert \right) _{\max }=N^{-1}\ln \left( \left(
\sqrt{3}+1\right) ^{2}/2\right) \approx 1.32/N$. Note that the largest
acceleration with which the soliton can be dragged by the moving defect, as
given by Eq. (\ref{amax}), does not depend on the sign of $\varepsilon $.

A different result is produced for dragging solitons by the local defect
moving with constant acceleration in the framework of the NLS equation with
the quintic, rather than cubic, nonlinearity (which may also be realized in
optical media \cite{Cid}),%
\begin{equation}
iu_{z}+(1/2)u_{xx}+|u|^{4}u=-\varepsilon \delta \left( x-az^{2}/2\right) u.
\label{quintic}
\end{equation}%
In the uniform space ($\varepsilon =0$), Eq. (\ref{quintic}) gives rise to
commonly known \textit{1D Townes solitons} \cite{Salerno}, $u=\left(
3k\right) ^{1/4}\sqrt{\mathrm{sech}\left( 2\sqrt{2k}x\right) }\exp \left(
ikz\right) $, with arbitrary propagation constant $k>0$. This soliton family
is degenerate, as its norm takes a single value, which does not depend on $k$%
, $N=\sqrt{3/2}\pi /2$, and the family is completely unstable against the
onset of the critical collapse \cite{Sulem-Sulem}. However, all the solitons
are \emph{stabilized} by the interaction with the quiescent attractive
defect ($a=0$, $\varepsilon >0$) \cite{Wang}. Then, the above consideration
can be developed for the solitons of Eq. (\ref{quintic}) pulled by the
defect with constant acceleration. In particular, the largest acceleration
which can be supported by the defect with given $\varepsilon >0$ is $%
\left\vert a_{\max }\right\vert =(4/\pi )\varepsilon k$, cf. Eq. (\ref{amax}%
).

The objective of this \textit{perspective} is to produce a brief summary of
results which predict possibilities of true SA motion of 1D and 2D solitons
in specific BEC\ models, one based on the spatially-periodic optical-lattice
(OL) potential, and another one making use of a binary BEC whose components
are resonantly coupled by a microwave (MW) field.

\textbf{Co-accelerating bound states of solitons with positive and negative
masses}. -- A two-component model which allows one to predict stable SA
bound states of solitons with positive and negative effective masses is
represented by a system of Gross-Pitaevskii (GP) equations for wave
functions $\phi $ and $\psi $ of the binary BEC, including the
spatially-periodic potential of the OL type in each equation, with strengths
$U_{1}$ and $U_{2}$ \cite{HS}:%
\begin{eqnarray}
i\phi _{t} &=&-(1/2)\phi _{xx}-\left[ g_{1}|\phi |^{2}+\gamma |\psi
|^{2}+U_{1}\cos \left( 2\pi x\right) +fx\right] \phi ,  \notag \\
i\psi _{t} &=&-(1/2)\psi _{xx}-\left[ \gamma |\phi |^{2}-g_{2}|\psi
|^{2}+U_{2}\cos \left( 2\pi x\right) +fx\right] \psi .  \label{fcos}
\end{eqnarray}%
Here, $g_{1}>0$ and $-g_{2}<0$ are coefficients of the self-interaction of
the components, implying that their signs are made opposite by means of the
Feshbach resonance applied to one of the components \cite{FR}, and $\gamma
>0 $ is the coefficient of the cross-attraction. The OL period in Eq. (\ref%
{fcos}) is set equal to $1$ by means of rescaling. The system also includes
a possibility to consider the action of gravity, with strength $f$, on both
components.

\textit{Analytical considerations}. To provide opposite signs of the
effective mass for solitons in components $\phi $ and $\psi $, it is natural
to consider the case when quasi-wavenumbers of wave functions $\phi $ and $%
\psi $ are set to be close, respectively, to the center and edge of the
first OL's Brillouin zone. The respective effective masses, calculated by
means of the known methods for the linear GP equation \cite{Pu,we,gapsol},
are, respectively,
\begin{equation}
M_{1}=\frac{2\pi ^{3}+U_{1}^{2}+\pi ^{2}\sqrt{4\pi ^{4}+2U_{1}^{2}}}{10\pi
^{4}+U_{1}^{2}-3\pi ^{2}\sqrt{4\pi ^{4}+2U_{1}^{2}}},  \label{meff1}
\end{equation}%
\begin{equation}
-M_{2}=U_{2}/\left( U_{2}-2\pi ^{2}\right)   \label{meff2}
\end{equation}%
($M_{2}$ is defined with sign minus, to focus on the relevant case of the
negative mass, which means $M_{2}>0$), and the wave functions themselves are
approximated by
\begin{equation}
\phi (x)=\Phi (x)\frac{1+2a\cos \left( 2\pi x\right) }{\sqrt{1+2a^{2}}},\psi
(x)=\sqrt{2}\Psi (x)\cos \left( \pi x\right) ,  \label{phi}
\end{equation}%
with $a\equiv \sqrt{\left( \pi ^{2}/U_{1}\right) ^{2}+1/2}-\pi ^{2}/U_{1}$,
where $\Phi (x)$ and $\Psi (x)$ are slowly varying (in comparison with $\cos
(2\pi x)$) envelopes. Actually, the negative mass is a characteristic
feature of gap solitons, generated by the interplay of the OL potential and
self-repulsion \cite{we}.

The substitution of expressions (\ref{phi}) in Eqs. (\ref{fcos}) leads, by
means of the averaging procedure, to equations governing the slow evolution
of the envelope amplitudes, which do not include the OL potential,
\begin{eqnarray}
i\frac{\partial \Phi }{\partial t} &=&-\frac{1}{2M_{1}}\frac{\partial
^{2}\Phi }{\partial x^{2}}-\left( G_{1}|\Phi |^{2}+\Gamma |\Psi
|^{2}+fx\right) \Phi ,  \label{PHI} \\
i\frac{\partial \Psi }{\partial t} &=&\frac{1}{2M_{2}}\frac{\partial
^{2}\Psi }{\partial x^{2}}-\left( \Gamma |\Phi |^{2}-G_{2}|\Psi
|^{2}+fx\right) \Psi ,  \label{PSI}
\end{eqnarray}%
with effective nonlinearity coefficients,
\begin{gather}
G_{1}=g_{1}\frac{1+12a^{2}+6a^{4}}{(1+2a^{2})^{2}},G_{2}=\frac{3}{2}g_{2},
\notag \\
\Gamma =\frac{1+2a^{2}+2a}{1+2a^{2}}>0.  \label{geff}
\end{gather}%
Sign minus is eliminated in front of the second derivative in Eq. (\ref{PSI}%
) according to the definition of the respective effective mass in Eq. (\ref%
{meff2}). Note that Eqs. (\ref{PHI}) and (\ref{PSI}) conserve the total
momentum of wave functions $\Phi $ and $\Psi $ (cf. Eq. (\ref{P})), in spite
of the possibility to produce SA states, as shown below.

It is obvious that Eqs. (\ref{PHI}) and (\ref{PSI}) may indeed feature
opposite signs of the effective masses, if $M_{1}$ and $M_{2}$ are both
positive (or both negative), and opposite signs of the self-interaction in
the two components, if $G_{1}$ and $G_{2}$ are both positive (or both
negative) too. These sign combinations open the way to the creation of
coupled pairs of solitons with opposite signs of their dynamical masses.

Using the Lagrangian structure of Eqs. (\ref{PHI}) and (\ref{PSI}), the
variational approximation (VA)\ can be applied to the system, using the
following ansatz for two-component solitons:
\begin{equation}
\left\{ \Phi ,\Psi \right\} =\sqrt{N_{1,2}}\left( 2\alpha _{1,2}/\pi \right)
^{1/4}\exp \left[ i\varphi _{1,2}-\alpha _{1,2}\left( x-\xi _{1,2}\right)
^{2}+ik_{1,2}\left( x-\xi _{1,2}\right) \right] ,  \label{ansatz}
\end{equation}%
with norms $N_{1,2}$, widths $\alpha _{1,2}^{-1/2}$, central coordinates $%
\xi _{1,2}$, momenta $k_{1,2}$, and phases $\varphi _{1,2}$. The VA\
procedure \cite{progress} leads to equations of motion for the coordinates
in the absence of gravity ($f=0$) \cite{HS}:
\begin{gather}
\frac{d^{2}\xi _{1,2}}{dt^{2}}=\frac{N_{2,1}}{M_{1,2}}\alpha \exp \left[ -%
\frac{2\alpha _{1}\alpha _{2}}{\alpha _{1}+\alpha _{2}}(\xi _{1}-\xi
_{2})^{2}\right] (\xi _{2}-\xi _{1}),  \label{xi1} \\
\alpha \equiv \left( 2\Gamma /\sqrt{\pi }\right) \left( 2\alpha _{1}\alpha
_{2}/(\alpha _{1}+\alpha _{2})\right) ^{3/2},\rho \equiv
N_{2}/M_{1}-N_{1}/M_{2}.  \label{alphabeta}
\end{gather}%
Further, in the linear approximation with respect to $\Delta \xi \equiv \xi
_{2}-\xi _{1}$ Eq. (\ref{xi1}) gives rise to the following equations for $%
\Delta \xi $ and the mean coordinate, $\Xi \equiv \left( \xi _{1}+\xi
_{2}\right) /2$:%
\begin{equation}
\frac{d^{2}\Delta \xi }{dt^{2}}=-\alpha \rho \Delta \xi ,\frac{d^{2}\Xi }{%
dt^{2}}=n\Delta \xi ,n\equiv \frac{\alpha }{2}\left( \frac{N_{2}}{M_{1}}+%
\frac{N_{1}}{M_{2}}\right) .~  \label{lin}
\end{equation}

In the case of $\rho >0$ (see Eq. (\ref{alphabeta})), the solution of Eqs. (%
\ref{lin}) predicts shuttle motion of the bound state of the solitons with
positive and negative masses, combined with its inner oscillations:%
\begin{equation}
\Delta \xi (t)=\Delta \xi _{0}\cos \left( \sqrt{\alpha \rho }t\right) ,~\Xi
(t)=\frac{n}{\alpha \rho }\Delta \xi _{0}\left[ 1-\cos \left( \sqrt{\alpha
\rho }t\right) \right] ,  \label{shuttle}
\end{equation}%
where $\Delta \xi _{0}$ is an arbitrary amplitude of the oscillations.
Examples of the shuttle motion produced by numerical simulations of Eqs. (%
\ref{PHI}) and (\ref{PSI}) are displayed below in Figs. \ref{fig1}(b) and %
\ref{fig2}(a).

The SA motion of the two-soliton bound state is predicted at $\rho =0$,
i.e., as it follows from Eq. (\ref{alphabeta}), for relation $%
N_{1}/N_{2}=M_{2}/M_{1}$ between the norms and effective masses of the two
components. This condition provides cancellation of the force driving the
evolution of $\Delta \xi $. In this case, Eqs. (\ref{shuttle}) demonstrate
that the separation $\Delta \xi _{0}$\ between centers of the components
remains constant, and the solitons move with identical accelerations, so that%
\textbf{\ }$\Xi (t)=(n/2)\Delta \xi _{0}t^{2}$\textbf{.} Examples of the SA
motion produced by simulations of Eqs. (\ref{PHI}) and (\ref{PSI}) are
presented below in Figs. \ref{fig1}(a) and \ref{fig2}(a). Finally, in the
case of $\rho <0$ Eqs. (\ref{lin}) predict the exponential growth of $\Delta
\xi (t)$, i.e., splitting of the two-soliton pair, which is also confirmed
by direct numerical solutions.

Further, the VA was extended to include the gravity terms in Eqs. (\ref{PHI}%
) and (\ref{PSI}). A straightforward analysis predicts the shift of the
point at which the SA state appears from $\rho =0$ at $f=0$ to $\rho
=-\left( M_{1}M_{2}\alpha \Delta \xi \right) ^{-1}\left( M_{1}+M_{2}\right)
\exp \left[ 2\alpha _{1}\alpha _{2}\left( \alpha _{1}+\alpha _{2}\right)
^{-1}\left( \Delta \xi \right) ^{2}\right] f$. Thus, the gravity may be used
to control the occurrence of the SA regime. The appropriate value of $f$ can
be tuned, in turn, by varying the angle, $\Theta $, between the vertical
axis and direction of the quasi-1D waveguide into which the BEC is loaded: $%
f=f_{\max }\cos \Theta $.

\textit{Numerical results}. Typical examples of the dynamics of the bound
states of solitons with effective positive and negative masses, produced by
simulations of Eqs. (\ref{PHI}) and (\ref{PSI}) with parameters and inputs%
\begin{equation}
M_{1}=1,G_{1}=0.9,\Gamma =0.1,G_{2}=0.1+M_{2}^{-1},  \label{parameters}
\end{equation}%
\begin{equation}
\Phi _{0}=\mathrm{sech~}x,\Psi _{0}=\mathrm{sech}\left( x-\Delta \xi \right)
,\Delta \xi =0.1  \label{inputs}
\end{equation}%
(i.e., the soliton norms are $N_{1}=N_{2}=2$) are displayed in Fig. \ref%
{fig1}. In particular, $M_{2}^{-1}=1$ and $M_{2}^{-1}=0.96$ in Figs. \ref%
{fig1}(a) and (b) correspond, respectively to $\rho =0$ and $\rho =0.08$
(see Eq. (\ref{alphabeta})), which, in agreement with the VA prediction,
produce the SA and shuttle motion, respectively. For values of $M_{2}^{-1}>1$%
, i.e., $\rho <0$, the simulations demonstrate splitting of the two-soliton
state, also as predicted by the VA (not shown here).
\begin{figure}[h]
\includegraphics[height=4.cm]{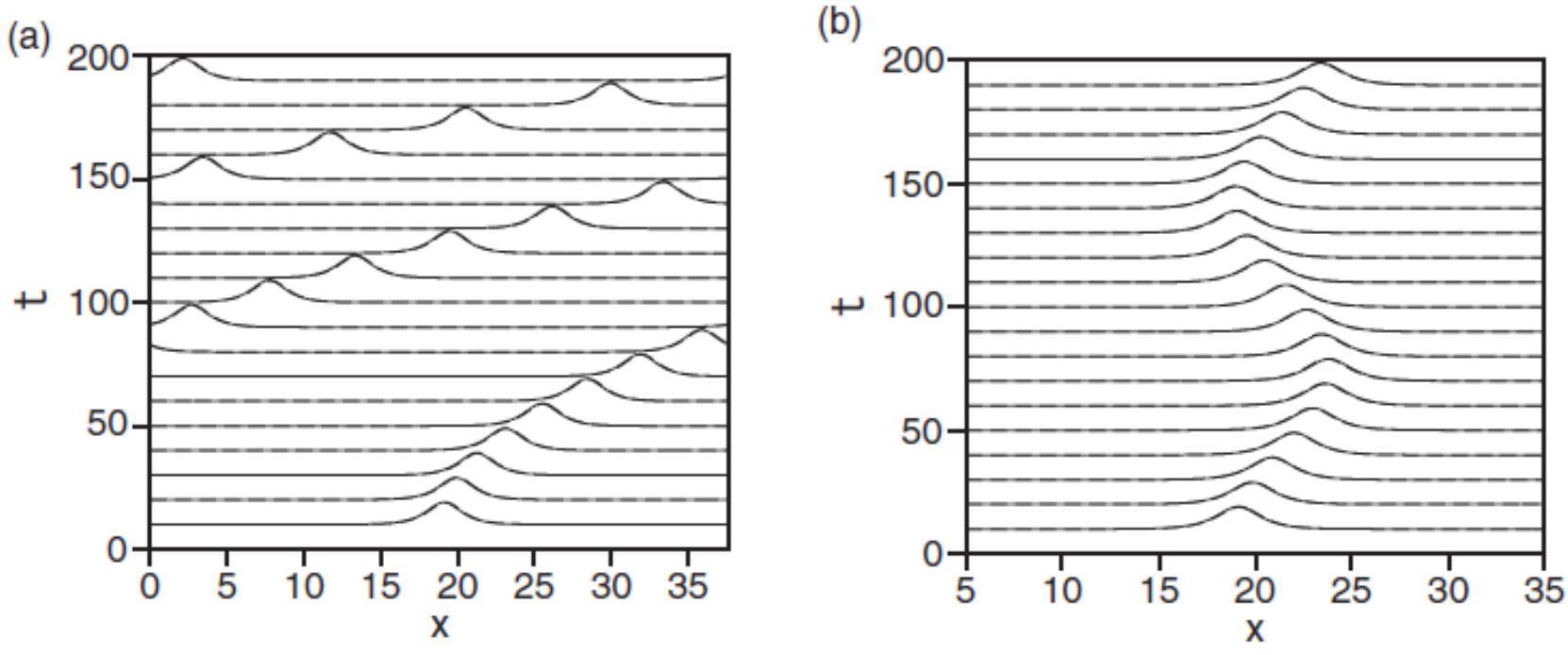}
\caption{The evolution of $\left\vert \Phi \left( x,t\right) \right\vert $
and $\left\vert \Psi \left( x,t\right) \right\vert $ (nearly overlapping
solid and dashed lines, respectively) produced by simulations of Eqs. (%
\protect\ref{PSI}) and (\protect\ref{PHI}) with input (\protect\ref{inputs})
and parameters (\protect\ref{parameters}), where $M_{2}^{-1}=1$ (a) and $%
M_{2}^{-1}=0.96$ (b) (as per Ref. \protect\cite{HS}).}
\label{fig1}
\end{figure}

Additional numerical results are presented in Fig. \ref{fig2}(a), which
shows the law of motion of central coordinates for $M_{2}^{-1}=1$, $0.98$,
and $0.92$, i.e., $\rho _{1;0.98;0.92}=0;0.04;0.16$, respectively. Note that
ratio $\sqrt{\rho _{0.92}/\rho _{0.98}}=2$ exactly corresponds to the ratio
of periods $T_{0.98}/T_{0.92}=2$, in agreement with Eq. (\ref{shuttle}).
Further, Fig. \ref{fig2}(b) compares the numerically measured acceleration
of the SA pair and its VA-predicted counterpart, as given by Eq. (\ref{xi1}%
). In the presence of the gravity in Eqs. (\ref{PSI}) and (\ref{PHI}), the
direct simulations also accurately corroborate the VA\ predictions \cite{HS}%
.
\begin{figure}[h]
\subfigure[]{\includegraphics[height=4.cm]{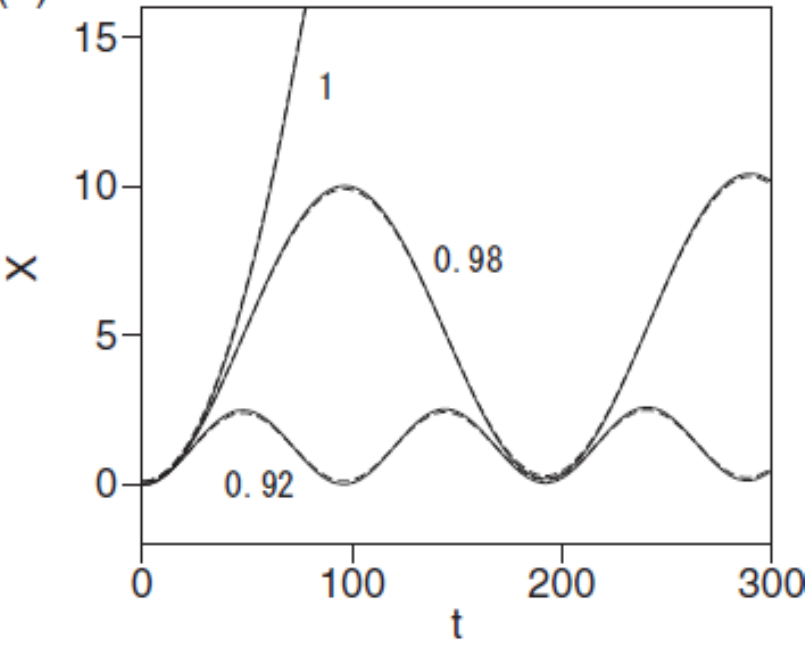}} \subfigure[]{%
\includegraphics[height=4.cm]{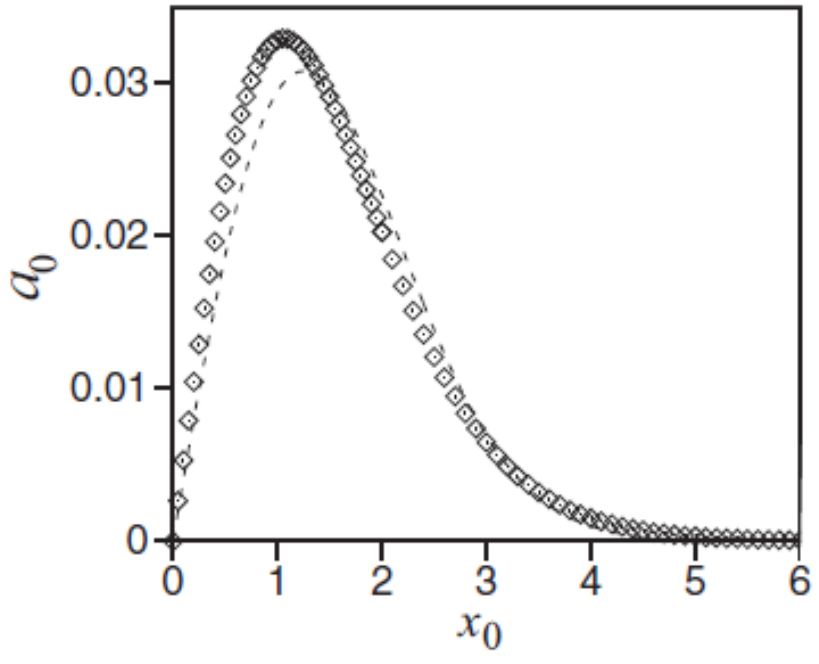}}
\caption{(a) Trajectories of centers of the $\Phi $ and $\Psi $ components
(solid and dashed lines, respectively), produced by the the same simulations
as in Fig. \protect\ref{fig1}, with $M_{2}^{-1}=1$, $0.92$, and $0.98$, as
indicated. (b) The acceleration of the SA bound state, $a_{0}$, vs. the
constant separation, $\Delta \protect\xi \equiv x_{0}$, between the bound
solitons with the positive and negative masses, in the case of $\protect\rho %
=0$ (see Eq. (\protect\ref{alphabeta}). Rhombuses: results of the
simulations of Eqs. (\protect\ref{PSI}) and (\protect\ref{PHI}) with input (%
\protect\ref{inputs}) and parameters (\protect\ref{parameters}), where $%
M_{2}^{-1}=1$. The dashed curve: the VA prediction produced by Eq. (\protect
\ref{xi1}) (as per Ref. \protect\cite{HS}).}
\label{fig2}
\end{figure}

\begin{figure}[h]
\subfigure[]{\includegraphics[height=4.cm]{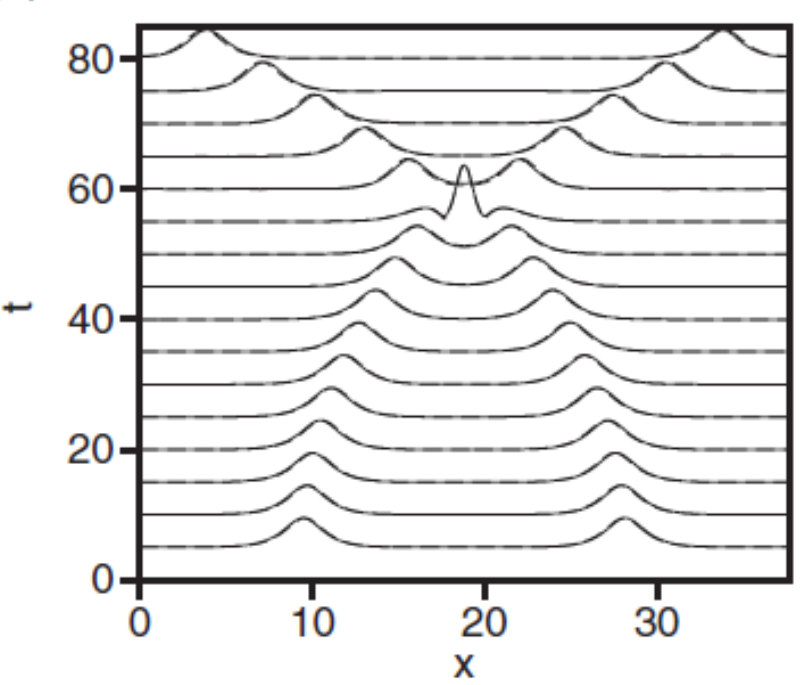}} \subfigure[]{%
\includegraphics[height=4.cm]{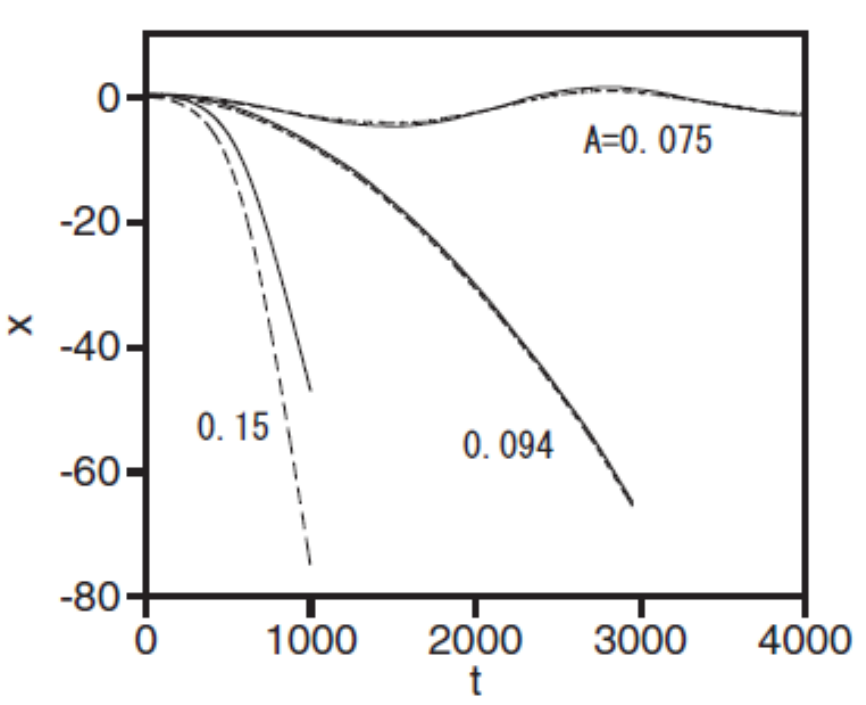}}
\caption{(a) The collision of the SA pair, shown in Figs. \protect\ref{fig1}%
(a) and \protect\ref{fig2}(a), and its mirror image moving with the opposite
self-acceleration. (b) Trajectories of motion of centers of components $%
\protect\phi $ and $\protect\psi $ (the continuous and dashed lines,
respectively) produced by simulations of the underlying system (\protect\ref%
{fcos}), see details in the text (as per Ref. \protect\cite{HS}).}
\label{fig3}
\end{figure}

The existence of robustly moving SA bound states suggest a possibility to
consider collisions between such modes moving with opposite accelerations
\cite{HS}. A typical example, \textit{viz}., collision between the same SA
state which is displayed above in Figs. \ref{fig1}(a) and \ref{fig2}(a) and
its mirror image is displayed in Fig. \ref{fig3}(a). It shows that the
colliding soliton pairs pass through each other, causing increase of the
separation between the bound solitons in each pair, from $\Delta \xi =0.10$
to $\Delta \xi \approx 0.17$. This, in turn, leads to the increase of the
co-acceleration, in agreement with Eq. (\ref{xi1}) and Fig. \ref{fig2}(b).

As concerns direct simulations, the results presented above were produced by
Eqs. (\ref{PSI}) and (\ref{PHI}), which were derived from the full system (%
\ref{fcos}) as equations for the slow evolution of amplitudes $\Phi $ and $%
\Psi $ in ansatz (\ref{phi}). Direct simulations of the full equations
produce similar results \cite{HS}. As an example, Fig. \ref{fig3}(b)
displays trajectories of centers of the two components, as obtained from
simulations of Eqs. (\ref{fcos}) with $U_{1}=0$, $U_{2}=8$, $f=0$ (no
gravity) and input given by Eq. (\ref{phi}), with $\Phi (x)$ and $\Psi (x)$
replaced by amplitudes $A=0.150$, $0.094$, $0.075$ and $B=0.150$. As seen in
the figure, the stable SA bound state is generated by $A=0.094$, while the
VA predicts, for the same input, the SA state at $A=0.102$. The
corresponding relative error $\approx $ \ $0.08$ demonstrates the accuracy
of the VA, in comparison with the simulations of the full underlying system (%
\ref{fcos}). The shuttle motion and splitting are observed at $A=0.075$ and $%
0.150$, respectively, in agreement with the VA. In Ref. \cite{Yulik},
similar simulations were performed for the system including, in addition to
the OL, the quadratic potential in each component. It produces opposite
forces acting on the positive- and negative-mass components, leading to
their splitting.

If the solitons are considered in a more general form than defined by Eq. (%
\ref{phi}), the formation of the SA bound states of solitons with positive
and negative masses is still possible (in particular, because gap solitons
are mobile in the general case \cite{we}). However, this possibility was not
explored in detail.

\textbf{Self-acceleration of vortex rings (VRs) in microwave-coupled binary
BEC.} -- The existence of stable 2D SA solitons, including VRs, in a binary
BEC\ formed by two atomic states $\phi _{\uparrow ,\downarrow }$, resonantly
coupled by the magnetic component $H$ of the MW field, was predicted in Ref.
\cite{Qin2}. The respective system of scaled GP equations for $\phi
_{\uparrow ,\downarrow }$ and the Poisson equation for $H$ are \cite{Qin1}%
\begin{eqnarray}
i\frac{\partial \phi _{\downarrow }}{\partial t} &=&\left( -\frac{1}{2}%
\nabla ^{2}+\eta \mathbf{-}\beta _{c}\left\vert \phi _{\uparrow }\right\vert
^{2}-\beta _{s}\left\vert \phi _{\downarrow }\right\vert ^{2}\right) \phi
_{\downarrow }-\pi H^{\ast }\phi _{\uparrow },  \notag \\
i\frac{\partial \phi _{\uparrow }}{\partial t} &=&\left( -\frac{1}{2}\nabla
^{2}-\eta \mathbf{-}\beta _{c}\left\vert \phi _{\downarrow }\right\vert
^{2}-\beta _{s}\left\vert \phi _{\uparrow }\right\vert ^{2}\right) \phi
_{\uparrow }-\pi H\phi _{\downarrow },  \label{down-up}
\end{eqnarray}%
\begin{equation}
\nabla ^{2}H=-\phi _{\downarrow }^{\ast }\phi _{\uparrow },  \label{Poisson}
\end{equation}%
where $\ast $ stands for the complex conjugate, $2\eta $\ is detuning of the
MW from the transition between the atomic states $\left\vert \uparrow
\right\rangle $\ and $\left\vert \downarrow \right\rangle $, $\beta _{c}$
and $\beta _{s}$\ being, respectively, strengths of the cross- and
self-interaction of the two components. Equations (\ref{down-up}) are
supplemented by the normalization condition, $\int \left( \left\vert \phi
_{\uparrow }\right\vert ^{2}+\left\vert \phi _{\downarrow }\right\vert
^{2}\right) d\mathbf{r}=1$. In the symmetric system with $\eta =0$, Eqs. (%
\ref{down-up}) coalesce into a single one for $\phi _{\downarrow }=\phi
_{\uparrow }\equiv \phi $, supplement by the accordingly simplified Poisson
equation for real $H$,%
\begin{equation}
i\frac{\partial \phi }{\partial t}=\left[ -\frac{1}{2}\nabla ^{2}-\beta
\left\vert \phi \right\vert ^{2}-\pi H\right] \phi ,\nabla ^{2}H=-|\phi
|^{2},  \label{symm}
\end{equation}%
with $\beta \equiv \beta _{c}+\beta _{s}$ and normalization $\int \left\vert
\phi (\mathbf{r})\right\vert ^{2}d\mathbf{r}=1/2$.

\begin{figure}[tbp]
\includegraphics[scale=0.45]{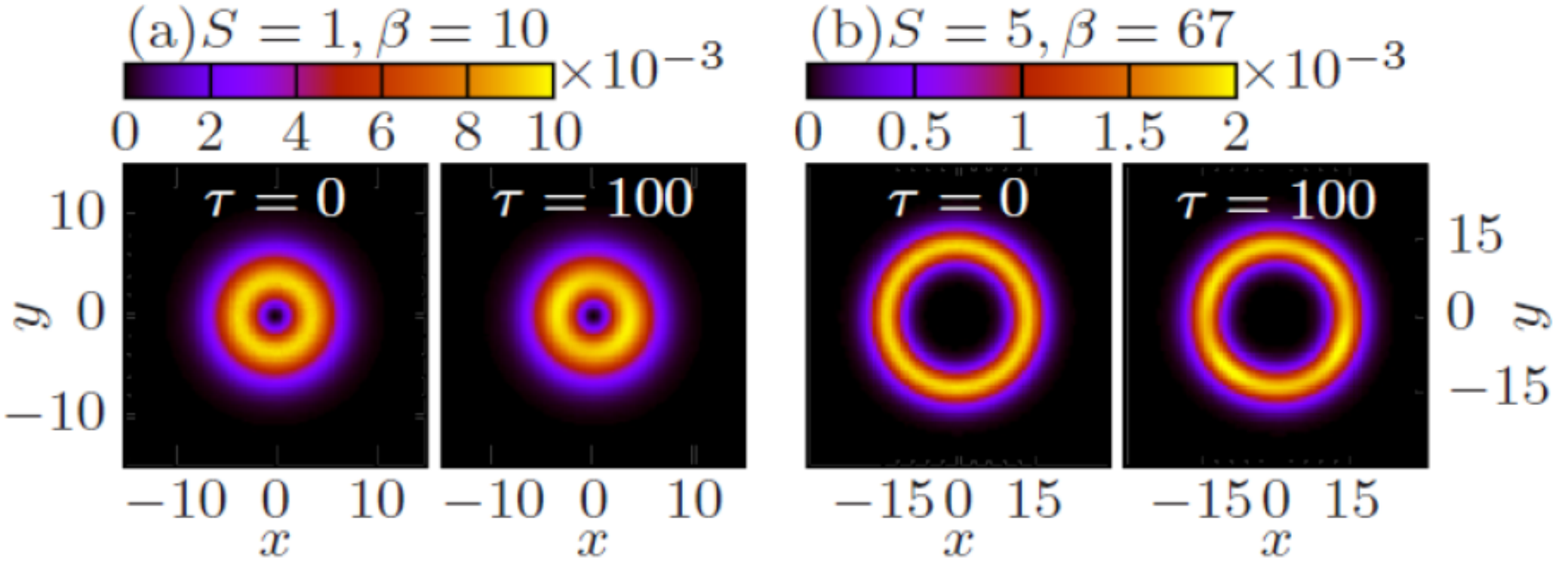}
\caption{Examples of the distribution of $\left\vert \protect\phi %
(r)\right\vert $ in stable VRs (vortex rings) with indicated values of $S$
and $\protect\beta $. The initial shapes of the VR ($\protect\tau =0$) are
compared to the outputs produced by simulations of Eq. (\protect\ref{symm})
at time $\protect\tau =100$ (as per Ref. \protect\cite{Qin1}).}
\label{fig4}
\end{figure}

Numerical solutions of Eq. (\ref{symm}) demonstrate that it gives rise to VR
solutions with chemical potential $\mu $ and vorticity $S=0,1,2,...$, in the
form of $\phi =\exp \left( -i\mu t+iS\theta \right) \Phi _{S}\left( r\right)
$, where\textbf{\ }$r$ and $\theta $ are the polar coordinates. The
solutions exist in the region of $\beta <\beta _{\max }(S)$, where $\beta
_{\max }(S=0)\approx 11.7$ corresponds to the commonly known value for
Townes solitons \cite{Sulem,Fibich}, and $\beta _{\max }(S\geq 1)$ are the
respective values for the Townes solitons with embedded vorticity \cite%
{Volkov}, which are well approximated by $\beta _{\max }(S)\approx 8\sqrt{3}%
\pi S$ \cite{Qin1}. In these intervals, the solitons with $S=0$ are
completely stable, while the VRs are stable in narrower regions, $\beta
<\beta _{\mathrm{st}}(S)$. An analytical approximation yields $\beta _{%
\mathrm{st}}(S)=2\sqrt{6}\pi S\approx \allowbreak 15.4S$, while an empirical
formula for the numerical results is $\beta _{\mathrm{st}}^{\mathrm{(num)}%
}(S)=15S-4$ \cite{Qin1}. Examples of stable VRs with $S=1$ and $5$ are
presented in Fig. \ref{fig4}. In intervals $\beta _{\mathrm{st}}(S)<\beta
\leq \beta _{\max }(S)$ the VRs are unstable against splitting into
necklace-shaped arrays of fragments. It is relevant to stress that the
growth of $\beta _{\mathrm{st}}^{\mathrm{(num)}}(S)$ $\symbol{126}~S$ for
the \textquotedblleft giant" VRs (ones with large values of $S$), produced
by Eq. (\ref{symm}), makes them much more robust modes than their
counterparts with smaller $S$. This feature is opposite to what was
previously found in those models which are able to produce stable VRs with $%
S>1$ \cite{Quiroga,Pego,Borovkova,Driben,Sudharshan,Reyna,Zhang,book}.

The finding reported in Ref. \cite{Qin2} is that Eqs. (\ref{symm}) are \emph{%
exactly invariant} with respect to a boost transformation, from the
quiescent reference frame to one which moves, in the 2D plane $\left(
x,y\right) $, with vectorial acceleration $\mathbf{a}=\left(
a_{x},a_{y}\right) $, combined with a constant velocity, $\mathbf{V}=\left(
V_{x},V_{y}\right) $. The coordinates, wave functions, and magnetic field in
the accelerating frame are%
\begin{equation}
\left\{ x^{\prime },y^{\prime }\right\} =\left\{ x.y\right\}
-V_{x,y}t-\left( a_{x,y}/2\right) t^{2},  \label{x'y'}
\end{equation}%
\begin{equation}
\phi ^{\prime }\left( x^{\prime },y^{\prime },t\right) =\phi \left(
x,y,t\right) \exp \left[ -i\left( a_{x}x+a_{y}y\right) t-i\left(
V_{x}x+V_{y}y\right) +i\chi (t)\right] ,  \label{phi'}
\end{equation}%
\begin{equation}
\chi (t)=(1/6)\left[ a_{x}^{-1}\left( V_{x}+a_{x}t\right)
^{3}+a_{y}^{-1}\left( V_{y}+a_{y}t\right) ^{3}\right] ,  \label{chi}
\end{equation}%
\begin{equation}
H^{\prime }\left( x^{\prime },y^{\prime },t\right) =H\left( x,y,t\right)
+\pi ^{-1}\left( a_{x}x+a_{y}y\right) .  \label{H'}
\end{equation}%
Actually, Eqs. (\ref{x'y'})-(\ref{H'}) are a generalization of the usual
Galilean boost for the accelerating reference frame. Note that the solution
of the 2D Poisson equation in system (\ref{symm}), with the source
represented by a quiescent soliton, has the standard asymptotic form far
from the region where the soliton is located (it is determined by the
Green's function for the 2D Laplacian):%
\begin{equation}
H(\mathbf{r})\approx -(1/2\pi )\left( \int |\phi (\mathbf{r}^{\prime })|^{2}d%
\mathbf{r}^{\prime }\right) \ln r.  \label{asympt}
\end{equation}%
The difference of the magnetic-field component (\ref{H'}) of the SA soliton
from its quiescent counterpart (\ref{asympt}) is the presence of the terms
linear in $x$ and $y$, which implies that the SA motion can be maintained by
the properly constructed background magnetic field. This field provides a
reservoir of the momentum which makes the self-acceleration possible.

According to Eq. (\ref{x'y'}), coordinates $\left( x_{c},y_{c}\right) $ of
the center of the stable 2D soliton moves as $x_{c}=V_{x}t+(1/2)a_{x}t^{2}$,~%
$y_{c}=V_{y}t+(1/2)a_{y}t^{2}$. This is a curvilinear trajectory in the 2D
plane: at small $t$, it is close to a straight line with slope $%
x/y=V_{x}/V_{y}$, while at $t\rightarrow \infty $ it is close to a line with
a different slope, $x/y=a_{x}/a_{y}$. In particular, in the case of $%
a_{x}=V_{y}=0$, the trajectory is a parabola:
\begin{equation}
y_{c}=\left( a_{y}/2V_{x}^{2}\right) x_{c}^{2}.  \label{parabola}
\end{equation}

\begin{figure}[h]
\subfigure[]{\includegraphics[scale=0.4]{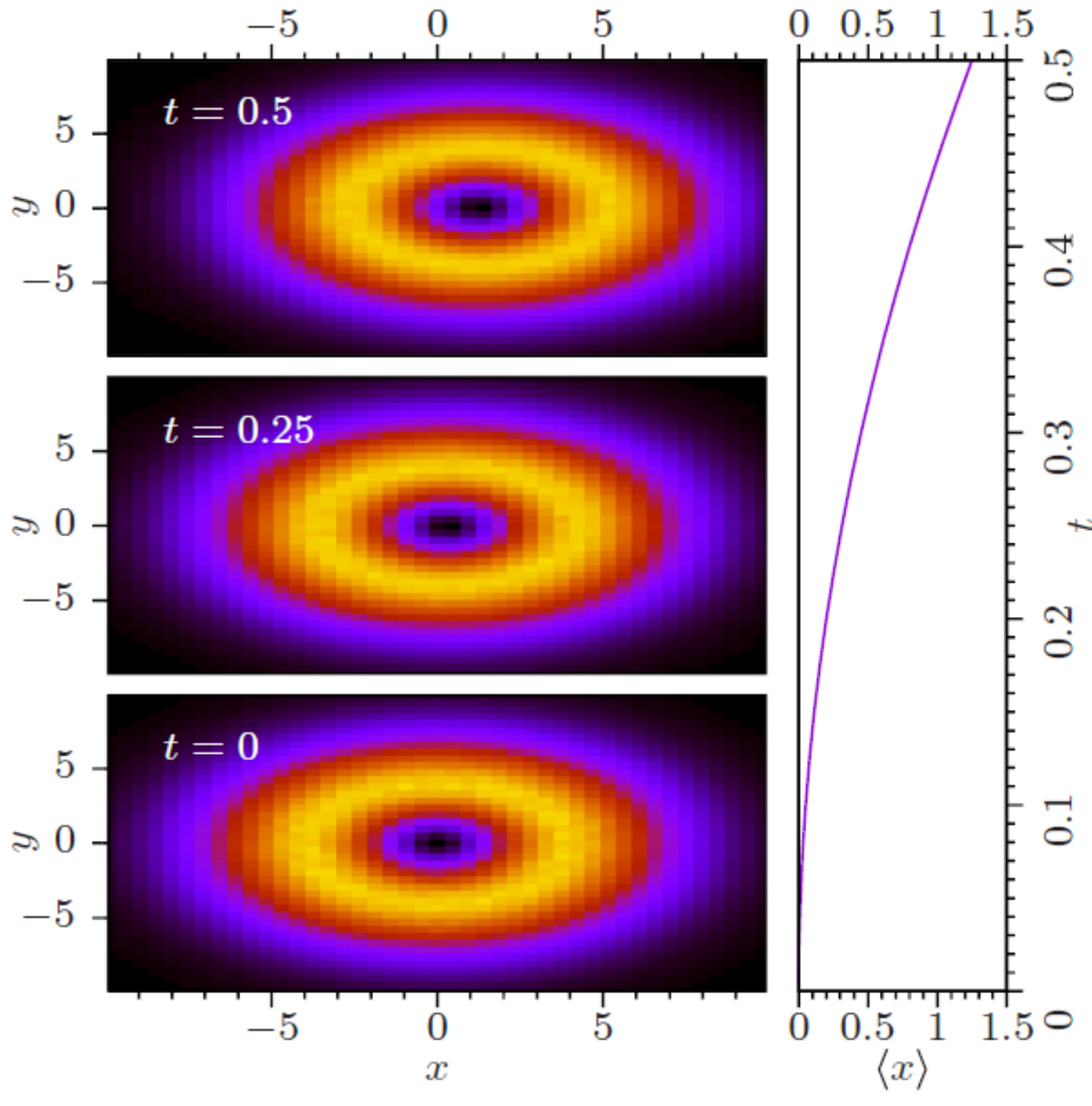}} \subfigure[]{%
\includegraphics[scale=0.4]{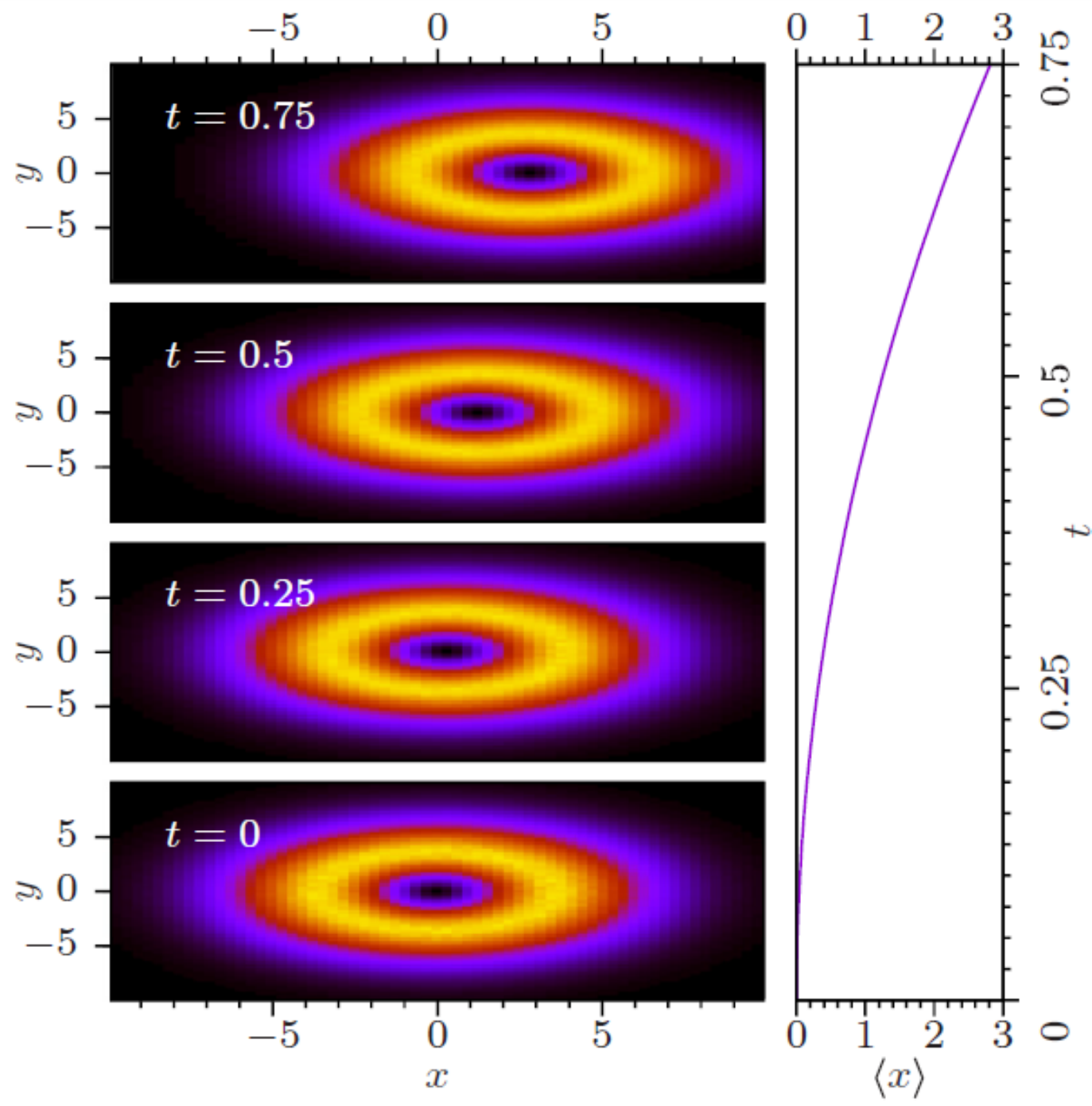}}
\caption{(a) The plot of $\left\vert \protect\phi \right\vert ^{2}$ for a
self-accelerating VR with $a_{x}=10$, $a_{y}=0$, $V_{x.y}=0$ (see Eq. (%
\protect\ref{x'y'})), produced by the numerical solution of Eq. (\protect\ref%
{symm}) with $\protect\beta =0$. (b) The same for $\protect\beta =10$. Right
panels in (a) and (b) show the time dependence of coordinate $\left\langle
x\right\rangle $ of the VR's center, cf. Eq. (\protect\ref{parabola}) (as
per \protect\cite{Qin2}).}
\label{fig5}
\end{figure}

The analytical results are corroborated by Fig. \ref{fig5}, which displays
stably moving VRs produced by simulations of Eq. (\ref{symm}). The numerical
solutions demonstrate exactly the same SA motion of the VRs as predicted by
Eq. (\ref{parabola}).
Essentially the same results were obtained for the SA motion of solitons
produced by the 1D version of Eq. (\ref{symm}), as well as for the 2D system
based on Eqs. (\ref{down-up}) and (\ref{Poisson}) including the spin-orbit
coupling between components $\phi _{\downarrow }$ and $\phi _{\uparrow }$
\cite{Qin2}.

\textbf{Conclusion}. -- This \textit{perspective} represents basic models
which make it possible to predict counterintuitive regimes of motion of
stable 1D and 2D solitons (including vortex rings) with SA
(self-acceleration). The corresponding models represent two-component BECs.
In one case, a pair of interacting 1D solitons with opposite signs of the
effective masses can be created in the binary BEC loaded in the OL potential
\cite{HS}. In that case, the opposite interaction forces, applied to the
solitons with opposite signs of the mass, produce the SA motion, similar to
what was previously predicted \cite{Peschel theory} and experimentally
realized \cite{Peschel-experiment} in a nonsoliton form. In the second case,
the system of GP equations for the matter-wave components, resonantly
coupled by the magnetic component of the MW (microwave) field, admits an
exact transition to the accelerating references frame, thus predicting
stable 2D and 1D SA solitons, including VRs \cite{Qin2}.

A challenging issue is a possibility of experimental realization of the
predicted SA states. The theory also has a potential for development. In
particular, as concerns the SA\ pairs of solitons with positive and negative
masses, it is relevant to extend the analysis for the 2D system.

\textbf{Acknowledgments -- }I thank Prof. Xiaojie Chen, EPL Deputy Editor,
for the invitation to write a \textit{perspective} article. This work was
supported, in part, by the Israel Science Foundation through grant No.
1286/17.


\begin{thebibliography}{99}
\bibitem{Belanger} Gagnon L. and Be1anger P. A., \textit{Opt. Lett}. \textbf{%
15} (1990) 466-468.

\bibitem{Parker} Parker D. F., Sophocleous C. and Radha C., \textit{J. Phys.
A: Math. Gen}. \textbf{35} (2002) 1283.

\bibitem{Berry} Berry M. V. and Balazs N. L., \textit{Am. J. Phys}. \textbf{%
47} (1979) 264-267.

\bibitem{Siviloglu-Christodoulides} Siviloglou G. A. and Christodoulides D.
N., \textit{Opt. Lett}. \textbf{32} (2007) 979-981.

\bibitem{Siviloglu et al 2007} Siviloglou G. A. \textit{et al.},
\textit{Phys. Rev. Lett}. \textbf{99} (2007) 213901.

\bibitem{electron waves} Voloch-Bloch N. \textit{et al.},
\textit{Nature} \textbf{494} (2013) 331-335.

\bibitem{Airy plasmonics} Minovich A. E. \textit{et al.}, \textit{Laser \&
Photonics Reviews} \textbf{8} (2013) 221-232.

\bibitem{kli} Efremidis N. K., Paltoglou V. and von Klitzing W., \textit{%
Phys. Rev. A} \textbf{87} (2013) 043637.

\bibitem{acc} Zhang P. \textit{et al.}, \textit{Nature Commun}. \textbf{5}
(2014) 4316.

\bibitem{gas-discharge} Clerici M. \textit{et al.}, \textit{Science Advances}
\textbf{1} (2015) 1400111.

\bibitem{Airy water} Fu S. \textit{et al.}, \textit{Phys. Rev. Lett}.
\textbf{115} (2015) 034501.

\bibitem{Ellenbogen2} Ellenbogen T. \textit{et al.},
\textit{Nature Phot}. \textbf{3} (2009) 395-398.

\bibitem{Airy nonlin Zhigang} Hu Y. \textit{et al.},
\textit{Opt. Lett.} \textbf{35}, 3952-3954 (2010).

\bibitem{Airy nonlin} Jia S. \textit{et al.},
\textit{Phys. Rev. Lett.} \textbf{104}, 253904 (2010).

\bibitem{Airy nonlin Segev} Kaminer I., Segev M., and Christodoulides D. N.,
\textit{Phys. Rev. Lett}. \textbf{106} (2011) 213903.

\bibitem{Lotti} Lotti A. \textit{et al.}, \textit{Phys. Rev. A} \textbf{84}
(2011) 021807.

\bibitem{Marom} Fattal Y., Rudnick A., and Marom D. M., \textit{Opt. Exp}.
\textbf{18} (2011) 17298-17307.

\bibitem{Thawatchai} Mayteevarunyoo T. and Malomed B. A., \textit{Opt. Lett}%
. \textbf{40} (2015) 4947-4950.

\bibitem{Peschel theory} Batz S. and Peschel U., \textit{Phys. Rev. Lett}.
\textbf{110} (2013) 193901.

\bibitem{Peschel-experiment} Wimmer M. \textit{et al.}, \textit{Nature Phys.}
\textbf{9} (2013) 780-784.

\bibitem{pert} Kivshar Yu. S. and Malomed B. A., \textit{Rev. Mod. Phys}.
\textbf{61} (1989) 763-915.

\bibitem{Cid} Reyna A. S. and C. B. de Ara\'{u}jo, \textit{Adv. Opt. Phot}.
\textbf{9} (2017) 720-774.

\bibitem{Salerno} Abdullaev F. Kh. and Salerno M., \textit{Phys. Rev. A}
\textbf{72} (2005) 033617.

\bibitem{Sulem-Sulem} Sulem C. and Sulem P.-L., \textit{The Nonlinear Schr%
\"{o}dinger Equation: Self-Focusing and Wave Collapse} (Springer, New York,
1999).

\bibitem{Wang} Wang L., Malomed B. A. and Yan Z., \textit{Phys. Rev. E}
\textbf{99}, 052206 (2019).

\bibitem{HS} Sakaguchi H. and Malomed B. A., Phys. Rev. E \textbf{99} (2019)
022216.

\bibitem{FR} Chin C. \textit{et al}., \textit{Rev. Mod. Phys}. \textbf{82}
(2010) 1225.

\bibitem{Pu} Pu H.\textit{\ et al}.,
Phys. Rev. A \textbf{67} (2003) 043605.

\bibitem{we} H. Sakaguchi and Malomed B. A., 
J. Phys. B \textbf{37}, 1443-1459 (2004).

\bibitem{gapsol} Brazhnyi V. A. and Konotop V. V., \textit{Mod. Phys. Lett. B%
} \textbf{18 }(2004) 627.

\bibitem{progress} Malomed B. A, Progr. Optics \textbf{43} (2002) 71-193.

\bibitem{Yulik} Bludov Yu. V. and Garc\'{\i}a-\~{N}ustes M. A., J. Phys. B:
At. Mol. Opt. Phys. \textbf{50} (2017) 135004.

\bibitem{Qin2} Qin J. \textit{et al.}, \textit{Phys. Rev. A} \textbf{99}
(2019) 023610.

\bibitem{Qin1} Qin J., Dong G., and Malomed B. A., \textit{Phys. Rev. A}
\textbf{94} (2016) 053611.

\bibitem{Sulem} Sulem C. and Sulem P.-L., \textit{The Nonlinear Schr\"{o}%
dinger Equation: Self-Focusing and Wave Collapse} (Springer, New York, 1999).

\bibitem{Fibich} Fibich G., \textit{The Nonlinear Schr\"{o}dinger Equation:
Singular Solutions and Optical Collapse} (Springer, Heidelberg, 2015).

\bibitem{Volkov} Kruglov V. I. \textit{et al.}, \textit{J. Phys. A: Math. Gen%
}. \textbf{21} (1988) 4381-4395.

\bibitem{Quiroga} Quiroga-Teixeiro M. and Michinel H., \textit{J. Opt. Soc.
Am. B} \textbf{14} (1997) 2004-2009 (1997).

\bibitem{Pego} Pego R. L. and Warchall H. A., \textit{J. Nonlinear Sci}.
\textbf{12} (2002) 347-394 (2002).

\bibitem{Borovkova} Borovkova O. V. \textit{et al.}, \textit{Phys. Rev. E}
\textbf{84} (2011) 035602(R).

\bibitem{Driben} Driben R. \textit{et al.}, \textit{Phys. Rev. Lett}.
\textbf{112} (2014) 020404.

\bibitem{Sudharshan} Sudharsan J. B. \textit{et al.}, \textit{Phys. Rev. A}
\textbf{92} (2015) 053601.

\bibitem{Reyna} Reyna A. S. \textit{et al.}, \textit{Phys. Rev. A} \textbf{93%
} (2016) 013840.

\bibitem{Zhang} Zhang H. \textit{et al.}, \textit{Opt. Lett}. \textbf{44}
(2019) 3098-3101.


\bibitem{book} Malomed B. A., \textit{Multidimensional Solitons} (AIP
Publishing, Melville, 2022).
\end{thebibliography}
\end{document}